\documentclass{article}

\newcommand{\bra}[1]{\big\langle#1\big\vert}
\newcommand{\ket}[1]{\big\vert#1\big\rangle}
\newcommand{\tr}{\text{Tr}}

\title{Environment-induced mixing processes in quantum walks}
\author{Lauri Lehman}
\date{}

\usepackage[pdftex]{graphicx}
\usepackage{amsmath,amssymb}

\begin{document}

\maketitle

\begin{center}
JARA Institute for Quantum Information,\\
Otto-Blumenthal-Stra{\ss}e, RWTH Aachen, 52074 Aachen, Germany\\
llehman@physik.rwth-aachen.de
\end{center}
\vspace{2ex}

\begin{abstract}
The mixing process of discrete-time quantum walks on one-dimensional
lattices is revisited in a setting where the walker is coupled
to an environment, and the time evolution of the walker and the
environment is unitary.
The mixing process is found to be incomplete, in the sense that the
walker does not approach the maximally mixed state indefinitely,
but the distance to the maximally mixed state saturates to some
finite value depending on the size of the environment.
The quantum speedup of mixing time is investigated numerically
as the size of the environment decreases from infinity to a
finite value.
The mixing process in this unitary setting can be explained by
interpreting it as an equilibration process in a closed quantum
system, where subsystems can exhibit equilibration even
when the entropy of the total system remains zero.
\end{abstract}
\vspace{2ex}

\section{Introduction}

Quantum walks have emerged in recent years as a useful theoretical tool to study
many fundamental aspects of quantum dynamics. The quantum walk model provides a
platform to study the effects of decoherence and disorder at the microscopical level,
in a simple setting that often allows the full analytical treatment of the problem.
The running times of numerical calculations also scale favourably in the size of the
system and in the running time of the walk, and large scale numerical calculations are
therefore possible.

The notion of mixing processes originates from random walks,
where the system's tendency to approach a completely random (``mixed")
state over a long time period is called the mixing process.
The completely mixed state at the end of the mixing process is
characterized by a uniform probability distribution over the nodes
of the graph, such that the walker can be found on any lattice site
with equal probability.
Since the final state is the same for any initial state, all the
information about the initial position of the walker is lost,
and the walk is said to have no memory (or to be Markovian).
In unitary quantum walks, the total state of the system is always pure and
the entropy is zero, so memory is preserved in the total wave function of
the system.
The system will explore all the possible configurations over time,
and shows no tendency to approach a particular state such as the
completely mixed state.
If the walker is subject to decoherence, the unitarity is lost and the system
evolves towards the completely mixed state.\cite{kendon2007}
Surprisingly, some amount of decoherence can even improve the
mixing rate in quantum systems.\cite{kt2003,richter2007}

The process of decoherence in a quantum system is attributed to the existence
of a coupling between the system and its environment.
When the system and the environment interact with each other, they become
entangled with each other, and a fully quantum mechanical treatment requires
that both systems are described by density matrices, probabilistic mixtures
of pure states.
In previous studies, the environment is usually described as a Markovian bath
that interacts with the system with some probability, and the dynamics
of the environment is completely neglected.
Such models do not take into account the correlations between noise events
which might arise when the environment has a memory.
The Markovian nature of the environment in these models implies that any
information about the initial state of the system is eventually lost, and
the system approaches the maximally mixed state asymptotically.
It is however interesting to ask how the memory of the environment affects
the dynamics of the mixing process.
In this article, the treatment of the environment is fully quantum mechanical,
and the emergence of the mixing process is studied as the size of
the environment increases.

The effects of decoherence in quantum walks have been studied extensively
in the past\cite{kendon2007}.
The first studies\cite{kt2003} concentrated on the mixing properties of the
time-averaged probability distribution, and studies on non-averaged
walks followed.\cite{km2008,richter2007}
Recently, the concepts of quantum stochastic walks\cite{wra2010}
and open quantum walks\cite{aps2012,apss2012} have been introduced,
generalizing the formalism of quantum walks under decoherence,
but mixing processes on finite graphs were not studied.
Hines and Stamp\cite{hs2011} have studied effects of non-Markovian
noise processes on continuous-time quantum walks,
and found that in the presence of a non-Markovian bath,
the propagation of the walker is ballistic.
When the environment is described quantum mechanically as a
spin bath, the mean-square displacement is reduced only by a constant
factor, and even strong coupling with the environment does not
induce diffusive behaviour in the long time limit, in stark contrast
with Markovian noise models.
Further similarities between the model introduced here and other
models in the literature are discussed in the next section.

Here, the mixing properties of quantum walks are studied numerically
using a discrete-time model with a finite number of sites.
The walk takes place on a one-dimensional chain with periodic boundary
conditions.
When the Hilbert space of the environment is finite, the mixing behaviour of
the walk was found to terminate after a certain number of time steps.
During the initial mixing process, the distance to the maximally mixed state
decreases exponentially, with a faster rate than the corresponding
classical random walk.
The mixing time of the walk was studied as a function of the size
of the environment, and the mixing time was found to approach the
classical value as the size of the environment increases.
After the state has converged to a certain distance from the maximally mixed
state, it saturates to an almost constant level, with only small fluctuations
around the mean value.
The average distance to the maximally mixed state in the long time limit
has a power-law dependency on the ratio of the sizes of the environment
and the lattice.
These results hold when the coupling between the system and the
environment is non-local, but studies on local environments show similar
properties.
The effect of non-commutativity of the environment matrices was also
studied in a local model, and increasing the degree of non-commutativity
was found to have similar effects as increasing the size of the environment.

\section{Quantum walk with an environment}

The transition of quantum states to classical states can be explained by
decoherence of quantum states, and decoherence itself can be explained by
interaction of the system with an environment.\cite{zurek2003}
To study how the existence of the environment induces mixing processes
in quantum walks, let us introduce a generic model for a quantum mechanical
environment.
The states of the environment are written
$\ket{\epsilon} = \sum_e \epsilon_e\ket{e}$
and belong to the Hilbert space
$\ket{\epsilon}\in\mathcal{H}_E$,
where the vectors $\ket{e}$ form an orthonormal basis for
$\mathcal{H}_E$ and the dimension of the space is denoted
dim$(\mathcal{H}_E) \equiv d_E$.
The assumptions about the structure of the environment are kept to a minimum,
in particular it is assumed that there are no symmetries associated with
the environment.

The Hilbert space of the usual discrete-time quantum walk consists of position
and coin.
The full quantum state of the walker with an environment at time $t$
can thus be written as
$\ket{\Psi(t)} = \sum_{s,c,e} \psi_{sce}(t)\ket{s}\ket{c}\ket{e}$,
where the vectors $\ket{s}\in\mathcal{H}_S$ and $\ket{c}\in\mathcal{H}_C$
form orthonormal bases for position and coin, respectively.
The coin space is two-dimensional and the dimension of the position
space is denoted dim$(\mathcal{H}_S) \equiv d_S$.
The full Hilbert space is thus
$\mathcal{H} = \mathcal{H}_S \otimes \mathcal{H}_C \otimes \mathcal{H}_E$,
and the time evolution is unitary in the full space.
The discrete time quantum walk evolves in discrete steps, with each step consisting
of a coin flip and a conditional translation in position space.\cite{abnvw2001,aakv2001}
The coin flip is represented by a unitary matrix $F$, given by the Hadamard matrix:
\begin{equation}
F\ =\ \frac{1}{\sqrt{2}} \begin{pmatrix}1&1\\1&-1\end{pmatrix}.
\end{equation}
The conditional translation operators \ $T_- \otimes P_0$ \ and \ 
$T_+ \otimes P_1$, \ with
$T_{\pm}\, =\, \sum_s \ket{s\pm1}_S \bra{s}$ and
$P_c\ =\ \ket{c}_C \bra{c}$,
project the coin state to either $\ket{0}$ or $\ket{1}$, accompanied by a translation to the
neighbouring site on the left or right, respectively.
To include the coupling between the system and the environment, some additional
dynamics must be introduced such that the system becomes entangled with
the environment.
For this purpose, introduce two unitary matrices $E_0$ and $E_1$ that
transform the state of the environment at each time step, in a conditional
way such that $E_0$ is applied for left moving components and $E_1$
for right moving components.
The time evolution for one time step of the walk is now given by the unitary
time evolution operator $U$:
\begin{equation}
\ket{\Psi(t+1)}\ =\ U \ket{\Psi(t)}
\end{equation}
\begin{equation}
U\ =\ T_- \otimes P_0F \otimes E_0\ +\ T_+ \otimes P_1F \otimes E_1
\end{equation}
with periodic boundary conditions assumed at the ends of the chain.
The number of sites $d_S$ is always chosen to be odd, so that the walker can
be found on any site at any given time step, and not just on even or odd sites.
In the beginning of the walk, the wave function is initialized
in the product state $\ket{\Psi(0)} = \ket{s_0}\ket{c_0}\ket{\epsilon_0}$,
with $\ket{s_0}$ localized with unit probability on one of the sites on the ring.
The initial state of the coin and the environment are not relevant for the
mixing process, as discussed below.
The final state at time step $t$ is obtained simply by multiplying the wave function
by $U$ iteratively: $\ket{\Psi(t)} = U^t \ket{\Psi(0)}$.
The main object of interest is the reduced density matrix of position space,
which is obtained by tracing over the coin and environment DOFs:
$\rho_S(t) = \tr_{CE}\, \big(\rho(t)\big)$, where the total state is always pure,
$\rho(t) = \ket{\Psi(t)}\bra{\Psi(t)}$.


The matrices $E_0$ and $E_1$ are assumed to be random unitary matrices
acting in the Hilbert space of the environment $\mathcal{H}_E$.
The most important property of these matrices is that they should not be
commuting, $\big[ E_0,E_1 \big] \neq 0$.
Only in this case does the environment have a non-trivial effect on the dynamics.
It can be shown that if the matrices
do commute, the coupling with the environment has no effect on the spatial
probability distribution, and only the non-diagonal values of the spatial
density matrix $\rho_S$ acquire a position-dependent complex phase.
The random unitary matrices $E_c$ can be constructed in a variety of ways.
Here we take an elementary approach by picking a random Hermitian
matrix $H_c$ with elements inside a fixed region around $0+0i$, and
exponentiating the matrix: $E_c = \exp(-i H_c)$.
The matrices $H_c$ can thus be viewed as the generating Hamiltonians for
the dynamics of the environment, with time unit $\Delta t = 1$.

Since the total wave function now includes the environment DOFs,
the evolution of the walker in position and coin space is no more unitary.
To obtain the state of the walker in position and coin space, the environment
DOFs must be traced out, and if the walker is entangled with
the environment, the resulting state is a mixed state.
The effect of the environment is thus that it induces a completely positive (CP)
map on the quantum walker, with the number of decoherence channels equal
to the dimension of the environment.
Denoting the reduced density matrix of position and coin by $\rho_{SC}(t)$,
the effect of the CP map $\mathcal{E}$ can be written formally as
\begin{equation}
\rho_{SC}(t)\; =\; \mathcal{E}\big(\rho_{SC}(0)\big)\; 
=\; \sum\limits_{e=1}^{d_E}\ X_e\, \rho_{SC}(0)\ X_e^\dag
\end{equation}
where the matrices $X_e$ are the Kraus generators of the CP map
acting in the space $\mathcal{H}_S\otimes\mathcal{H}_C$
and satisfying $\sum_e X_e^\dagger X_e = \mathbb{I}_{SC}$.
Their expressions can be given as
\begin{equation}
X_e\; =\; \Big(\mathbb{I}_{SC}\otimes\bra{e} \Big)\ U^t\ 
\Big( \mathbb{I}_{SC}\otimes\ket{\epsilon_0} \Big)
\end{equation}
where the vectors $\ket{e}$ form an orthonormal basis for $\mathcal{H}_E$,
$\mathbb{I}_{SC}$ is the identity transformation in space
$\mathcal{H}_S\otimes\mathcal{H}_C$,
and $\ket{\epsilon_0}$ is the initial state of the environment.
Note that the CP map $\mathcal{E}$ describes the time evolution from the
initial state to the final state at time $t$.
This is in contrast to previous studies of mixing
processes,\cite{km2008,ks2005} where the CP map describes the time
evolution from time step $t$ to $t+1$.
The previous studies assume implicitly that the noise acts in the same way
at each time step, neglecting the time correlations of noise events which
arise when the environment has a memory.

The quantum walk model with an environment is similar in spirit to quantum walks
with multiple coins, first studied by Brun \emph{et al}.\cite{bca2003-1}
In these models the coin is still two-dimensional and the total evolution is unitary,
but the coins are picked from a finite set at each time step and are reused cyclically,
and the changing of the coins can be thought to be done by the environment.
The variance of quantum walks with multiple coins behaves initially like the
classical random walk, but changes to quadratic behaviour of the standard quantum
walk after the initial classical period.
Numerical calculations show that the model introduced in this section behaves
in a similar way when considered on an infinite line.
Increasing the number of coins in the model of Brun \emph{et al} increases
the length of the classical period in the beginning;
similarly increasing the size of the environment $d_E$ increases the duration
of classical propagation in the beginning.

The quantum walk with an environment also bears resemblance to quantum walks
with non-Abelian anyons,\cite{bekptw2010,lzbpw2011} where one mobile anyon moves
in a chain of stationary anyons, braiding with the stationary anyons as it hops
between the lattice sites. There the anyons possess a collective fusion Hilbert
space, and braids between the anyons induce non-trivial transformations in the
fusion space. While the fusion space can also be viewed as a kind of environment,
the quantum walks with anyons were only studied on infinite chains, and the
mixing processes in finite chains were not considered.

In the following, the total system is divided into the system $S$ and its bath $B$,
where the system consists of the position DOFs and the bath consists of the coin and
environment DOFs together. The total Hilbert space is thus decomposed as
$\mathcal{H} = \mathcal{H}_S \otimes \mathcal{H}_B$ with
$\mathcal{H}_B = \mathcal{H}_C \otimes \mathcal{H}_E$, and the dimension of the
bath is $d_B = 2 d_E$.
This notation makes explicit the fact that the coin and the
environment together can be viewed as a big coin with enlarged dimension.

Although this model can be viewed as a quantum walk with redundant coin
DOF's, the mixing behaviour is specific to this model.
A quantum walk with a multidimensional coin, such as a discrete Fourier
transform or Grover coin, with coin projection operators $P_0$ and $P_1$
defined as projections to two orthogonal subspaces, does not
exhibit mixing, except in the average sense, like the usual quantum walk
with a 2-dimensional coin.

\section{Mixing time}

In the classical random walk, the state of the system evolves always towards
the completely mixed state, if there is no memory in the system.
In the unitary quantum walk, the coin can be thought to act as a memory
of the system, and the probability distribution does not approach any
particular state (although the time-averaged probability distribution can
be shown to approach a certain state\cite{aakv2001}).
If the quantum walk is subject to to decoherence acting via random noise
events, the probability distribution is known to approach the completely
mixed state.\cite{km2008,richter2007}
Here we assess a different scenario where the decoherence is caused
by the presence of an environment.
Now the total wave function of the system evolves unitarily, the information
about the state of the environment is kept for the whole duration of the
walk, and the noise events at different time steps are correlated with
each other.
As discussed below, the mixing process is qualitatively different in this
case.
When the size of the environment is finite, the mixing process terminates
after a certain time.
Once the walker has converged within a certain distance from the completely
mixed state, the distance does not decrease anymore, and the walker stays
at almost constant radius from the completely mixed state.

For quantum systems, the completely mixed state is a diagonal density matrix,
denoted $\omega$, such that all the diagonal values are equal to each other:
$(\omega)_{s,s'} = \delta_{s,s'} / d_S$.
The degree of mixing (in position space) can be quantified by the distance to the
maximally mixed state $D_{\omega}(t)$, given by
\begin{equation} \label{yht:etaisyys}
D_{\omega}(t) = D(\rho_S(t),\omega)
\end{equation}
where the trace distance between two density matrices is defined as
\begin{equation}
D(\rho_1,\rho_2)\ =\ \frac{1}{2} \tr \sqrt{(\rho_1-\rho_2)^2}.
\end{equation}
During the mixing process, the distance $D_{\omega}(t)$ decreases as the
state approaches the mixed state $\omega$.
In a random walk or in a quantum walk subject to random noise events,
the walker approaches the state $\omega$ asymptotically and
$D_{\omega}(t) \rightarrow 0$ as $t\rightarrow\infty$.

The time evolution of the reduced density matrix $\rho_S(t)$ was studied numerically
for various sizes of the chain and the bath, and also for various initial states, coin
matrices $F$ and environment matrices $E_c$.
Generally, the time evolution can be divided into two regions, the initial relaxation regime and
the stabilized regime.
In the initial stage of the walk, the system is mixing (or ``equilibrating") and $D_{\omega}$
decreases exponentially in time.
When $D_{\omega}$ has reached a certain value, the mixing process ends, and $D_{\omega}$
stays almost constant in time.
The remainder of this section concentrates on the analysis of the initial
mixing process, while the stabilized regime is treated in the next section.

\begin{figure}
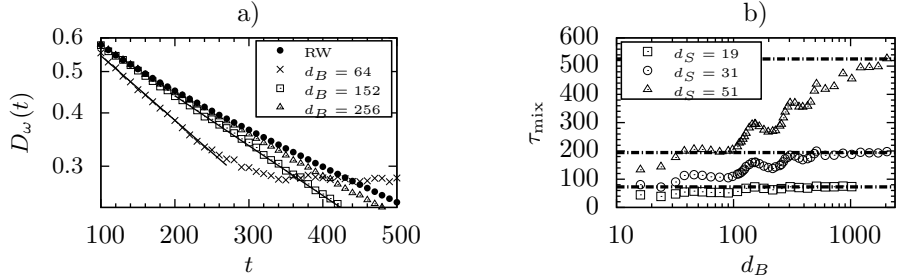

\begin{center}
\input{./etaisyys}
\hspace{3em}
\input{./sekoitusaika}
\end{center}
\caption{a) Typical time evolution of the distance $D_{\omega}(t)$ during the
initial relaxation period for different choices of the bath dimension $d_B$
(no averaging).
`RW' stands for classical random walk with equal hopping probabilities.
The total number of sites is $d_S = 51$.
Note that the scale on the y-axis is logarithmic.
The black lines indicate the best fit to an exponential function.
b) Quench-averaged mixing time as a function of the bath dimension $d_B$.
Each point is averaged over 30 samples, and the error bars (not shown)
are almost completely covered by the point marker.
The horizontal lines show the mixing time of the classical random walk
for each lattice size.}
\label{kuva:etaisyys}
\end{figure}

Typical results for the time evolution of the distance $D_{\omega}(t)$ are shown in
Fig. \ref{kuva:etaisyys} a), along with the results for the classical unbiased random walk.
These results are obtained for a specific choice of parameters, without any
averaging over initial states, environment matrices, or time.
Already at modest size of the environment $d_B$\,=\,256, the evolution is
close to the classical behaviour.
As the size of the environment decreases, the slope becomes slightly steeper,
indicating that the mixing process becomes faster.
This can be viewed as a quantum speedup of the mixing process.
Also visible at $d_B$\,=\,64 is the offset of the mixing process:
as the distance reaches the value of $D_{\omega}(t) \approx 0.28$, it
saturates to an almost constant level.
This also happens when the size of the bath is larger, but the threshold of
distance where this happens is smaller when $d_B$ gets larger.

Similar behaviour is also observed for the von Neumann entropy,
$H(\rho_S) = -\tr (\rho_S \ln \rho_S)$ (not plotted here).
In the initial stage of the walk, the entropy $H$ increases, but after
reaching a certain value it saturates to an almost constant level.
When the size of the environment increases, the saturated value of
$H$ approaches the maximal value $H_{\textrm{max}} = \ln d_S$,
which is the asymptotic value for a classical random walk.

\vspace{1em}
\vspace{1em}

Since the quantum walk with an environment does not approach the mixed
state $\omega$ indefinitely, the usual notion of mixing time is problematic,
because the state might never converge within a predetermined distance from
$\omega$.
A more suitable measure of mixing time is given by the slope of the
exponential curve:
\begin{equation}
D_{\omega}(t) \propto e^{-\frac{t}{\tau_{\textrm{mix}}}}.
\end{equation}
Defined in this way, the mixing time determines the time scale at which the distance  $D_{\omega}$
decreases by a factor of $1/e$.
The mixing time can then be calculated from an exponential fit for the
relevant time region $t_1 < t < t_2$.
Two examples of fitted curves are drawn as black lines in
Fig. \ref{kuva:etaisyys} a).

To understand how the mixing time changes as a function of space and bath dimensions,
the time evolution of the system was studied numerically for various configurations.
The results for the mixing time $\tau_{\textrm{mix}}$ are shown in
Fig. \ref{kuva:etaisyys} b).
Each point represents a quenched average over the environment: the walk was run with
different environment matrices, but keeping the same matrices during each run
(with a fixed initial state and coin).
When the size of the bath grows, the dynamics starts to resemble the classical random walk,
and the mixing time converges to the classical value given by the horizontal lines for each
system size.
In the intermediate region $d_B/d_S \approx 5$, the mixing time scales roughly
logarithmically in the size of the bath, although the numerical data are not
conclusive enough to decide between logarithmic and power-law dependence.
It can also be seen that the scaling function of the mixing time includes an
additional periodic component, causing the oscillations which can be seen
particularly at large system sizes.

The behaviour of the mixing time in Fig. \ref{kuva:etaisyys} b) can be viewed as
a transition between quantum and classical properties of the walk, with the
classical limit approached as {$d_B$}{$\rightarrow$}{$\infty$}.
The transition is rather smooth, so that increasing the size of the bath
by a small amount means that the dynamics is also a little bit closer to
the classical dynamics, although the oscillatory behaviour causes some
deviations.
Approximately logarithmic scaling of the mixing time implies that there is a large
region of intermediate bath sizes where there is some amount of quantum speedup.
On the other hand, logarithmic scaling would mean that decreasing the size of the
bath does not reduce the mixing time efficiently.
Indeed, as seen in Fig. \ref{kuva:etaisyys} a), decreasing the size of the
environment from infinity down to {$d_B$}\,=\,256 changes the slope by a very
modest amount.
Generally the speedup is more dramatic when the size of the lattice is large.
Comparing the speedup when the size of the environment decreases from
the classical value ({$d_B$}{$\rightarrow$}{$\infty$}) to
{$d_B$}\,{$\approx$}\,10 in Fig. \ref{kuva:etaisyys} b), the mixing time
is approximately halved when {$d_S$}\,=\,31, while for {$d_S$}\,=\,51
it reduces to one quarter of the classical value.

The time evolution of the walk is almost insensitive to the initial state
of the bath, and the changes for different initial states are of the same
order as for different environment matrices (with fixed dimension).
In contrast, the changes for different initial states of position can be drastic.
Starting from a non-localized state has essentially the same effect as increasing
the size of the bath, and the distance $D_{\omega}$ saturates to a higher level
at long time scales when the walker starts from a superposition of several
sites (only pure initial states were considered).
Also the choice of the coin can affect the dynamics.
While the long-time state is very similar for different choices of the coin,
the speed of the initial mixing process can vary for different coins,
the Hadamard coin being generally slower than other choices.

\section{Stabilized dynamics}

As seen in Fig. \ref{kuva:etaisyys} a) for {$d_B$}\,=\,64, the distance
$D_{\omega}$ stops to decrease after it has reached a certain value,
and remains at almost constant level from then on, with only small
fluctuations around the mean value.
When the size of the bath increases, the average distance becomes smaller and
the walker stays closer to the maximally mixed state $\omega$.
It is thus natural to ask how the average distance changes as a function of the bath size $d_B$.

\begin{figure}
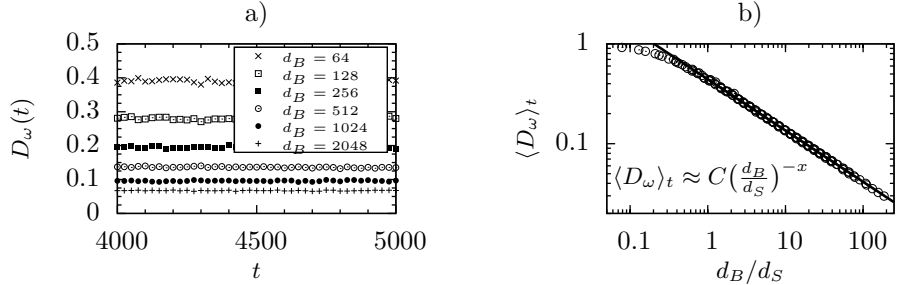

\begin{center}
\setlength{\unitlength}{.5\textwidth}
\input{./etaisyys-stabil}
\hspace{3em}
\input{./keskiarvo}
\end{center}
\caption{a) Distance to state $\omega$ after the termination of the
initial mixing process.
Depending on the bath size, distance $D_{\omega}$ saturates to a
certain average level.
The lattice size is $d_S$\,=\,51.
b) Average distance in long-time dynamics as a function of $d_B/d_S$.
The lattice sizes are $d_S = \{11, 19, 31, 51\}$.
See the main text for numerical estimates of the parameters $C$ and $x$.
}
\label{kuva:keskiarvo}
\end{figure}

The stabilized regime begins when the distance has saturated close
to the long-time average.
As seen in Fig. \ref{kuva:keskiarvo} a), fluctuations around the mean
value are small when $t$ is large, and the average distance
shows a clear dependency on the size of the bath.
To understand how the average distance depends on the bath size,
define the long time average of $D_{\omega}$ as
\begin{equation}
\big\langle D_{\omega} \big\rangle_t
= \frac{\sum_{t' = t_0}^t D_{\omega}(t')}{t-t_0+1}
\end{equation}
where $t_0$ is large enough so that $D_{\omega}$ is close to the long-time
average.
The time average $\big\langle D_{\omega} \big\rangle_t$ is plotted in
Fig. \ref{kuva:keskiarvo} b) for different choices of $d_S$ and $d_B$.
When plotted against the ratio $d_B/d_S$, the points for different lattice
sizes fall on the same line, showing that the long-time average depends
only on the ratio $d_B/d_S$.

\vspace{1em}
\vspace{1em}

The shape of the function in Fig. \ref{kuva:keskiarvo} b) suggests
that the average distance has a power law dependency on
the ratio $d_B/d_S$:
\begin{equation}
\langle D_{\omega} \rangle_t\ \approx\ C\,
\Big(\frac{d_B}{d_S}\Big)^{-x}
\end{equation}
when $d_B > d_S$. The numerical coefficients obtained by fitting are
$C = 0.4411 \pm 0.0013$ and $x = 0.5133 \pm 0.0013$, and the corresponding
function is drawn as a black line in Fig. \ref{kuva:keskiarvo} b).


The average distance is not the only property of stabilized dynamics which
is sensitive to the size of the bath.
Also the fluctuations around the average value change with $d_B$, and the
variance of the fluctuations becomes smaller as $d_B$ increases.
The effect of the environment is therefore that it drives the state of the system
to a small subset of all possible states, with the distance to the maximally
mixed state restricted to a certain interval.
When the size of the environment increases, the interval of allowed values
becomes smaller and the system is driven to a more restricted subset of
all states.

Eventually, the recurrence effect should bring the state close to its
initial state, and the distance to the maximally mixed state should increase.
The recurrence time in this system appears to be quite long however,
and the system stayed close to its long-time average for all cases studied.
In a few very long calculations, no recurrence was observed during 100 000
time steps of time evolution.

\section{Local environments}

In the previous description of the environment, it was assumed that the
environment acts in the same way for every position of the chain.
Thus the description of the environment is non-local, as the walker
sees the same environment at every position.
In a more realistic model, each site should have their own local environment,
such that the walker interacts with the environment only when it is
located in the respective position.
The mixing process was also studied by considering such local environments.

A simple model for a local environment can be constructed by attaching
an ancillary qubit to each lattice site.
When the walker occupies a given site, it interacts only with the qubit
attached to that site.
The environment consists of $d_S$ qubits, with dimension
$d_E = 2^{d_S}$, and the environment matrices carry also a position label.
The time evolution operator $U$ is now given by the expression
\begin{equation}
U\ =\ \sum\limits_s \Big(T_-^s \otimes P_0F \otimes E_0^s\
+\ T_+^s \otimes P_1F \otimes E_1^s \Big)
\end{equation}
where the site-dependent translation operator is written as
$T_{\pm}^s = |s\pm1\rangle_S \langle s|$, and the matrix $E_c^s$ acts
on qubit $s$ by a $2\times2$ matrix and trivially on other qubits.
Each qubit rotation can be parametrized using two real variables $\theta$
and $\gamma$. The environment matrices can be written as
$E_c^s =  \mathbb{I}^{\otimes (s-1)} \otimes G_c \otimes 
\mathbb{I}^{\otimes (d_S-s)}$, where the unitary matrix $G_c$ is given by
\begin{equation} \label{yht:ympmatrpaikall}
G_c = \begin{pmatrix} \cos\theta_c & -e^{-i\phi_c}\sin\theta_c\\
e^{i\phi_c}\sin\theta_c & \cos\theta_c \end{pmatrix}.
\end{equation}
The environment is assumed to be homogeneous and the same matrices
$G_c$ are applied at every site.

It was mentioned above that if the matrices $E_0$ and $E_1$ commute,
the environment has no effect on the probability distribution.
The extent to which these matrices fail to commute can thus be seen as a
measure of how strongly the system and the environment are coupled,
and intuitively the growth rate of entanglement is higher when the matrices
are less commuting.
To investigate how the mixing process is affected by the non-commutativity,
define the parameter $\gamma$ as the matrix norm of the commutator:
\begin{equation} \label{yht:gamma}
\gamma\ =\ \big\|\, \big[ G_0,G_1 \big]\, \big\|.
\end{equation}
Parametrization of the matrices according to Eq.
(\ref{yht:ympmatrpaikall}) ensures that the parameter $\gamma$
can be varied in a systematic way.

\begin{figure}
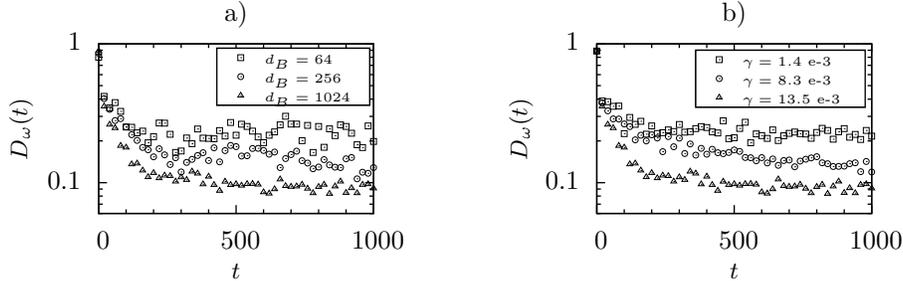

\begin{center}
\setlength{\unitlength}{.5\textwidth}
\input{./etaisyys-paikall-ulott}
\hspace{3em}
\input{./etaisyys-paikall-gamma}
\end{center}
\caption{
a) Time evolution of $D_{\omega}$ in a local environment for different
bath sizes.
The bath sizes are $d_B = \{64,256,1024\}$ corresponding to lattice
sizes $d_S = \{5,7,9\}$, respectively.
The environment matrices $E_c$ were chosen such that $\gamma=13.5$ e-3. 
b) Time evolution of $D_{\omega}$ in a local environment for different
values of $\gamma$.
The parameter $\gamma$ quantifies the noncommutativity of the environment
matrices, see Eq. (\ref{yht:gamma}).
The lattice size is $d_S = 9$ and the bath dimension is $d_B = 1024$.}
\label{kuva:etaisyys-paikall}
\end{figure}

The time evolution of $D_{\omega}$ in the local model is plotted in
Fig. \ref{kuva:etaisyys-paikall}.
The effect of increasing the parameter $\gamma$ in
Fig. \ref{kuva:etaisyys-paikall} b) is similar to increasing the
size of the environment in Fig. \ref{kuva:etaisyys-paikall} a):
the system saturates to a state which is closer to
the mixed state $\omega$.
The time evolution during the initial mixing process is too jumpy
to determine how the mixing time changes as a function of $\gamma$.
The large distance fluctuations in Fig. \ref{kuva:etaisyys-paikall} are probably
just a finite size effect of the lattice size, because similar fluctuations
are also seen in the non-local model for the corresponding sizes of the
lattice and the environment.
Unfortunately the dimension of the environment grows exponentially in the
number of sites, and numerical calculations become difficult for
larger systems.

\vspace{1em}
\vspace{1em}

Intuitively one might expect that the parameter $\gamma$ would act only
to slow down the mixing process, but in the long time limit
$D_{\omega}$ would saturate to the same value for all values of the
parameter $\gamma$.
The differences were however found to persist for very long times.
Even after 50 000 time steps, the mean distances have not changed significantly
from Fig. \ref{kuva:etaisyys-paikall} b).

The local model exhibits essentially the same properties as the
non-local model.
When the size of the environment is larger, the system saturates closer to
the maximally mixed state.
Changing the initial state of the bath affects only the details of time
evolution, not the asymptotic state, but starting from an initial
state which is spread out in position space causes the walker to
remain in state with a higher distance to $\omega$.
Comparing the local and non-local models with equally sized systems
and baths, the non-local model is more ``classical'', in the sense that
the non-local walk saturates closer to $\omega$ than the local walk.

\section{Conclusions}

The emergence of the mixing process in quantum walks was studied in a
finite-dimensional setting, where the walker becomes entangled with an
environment, and the system-environment dynamics is unitary.
The time evolution was found to be divided in two distinct regimes.
In the initial regime of time evolution, the distance to the maximally mixed
state decreases exponentially, and the rate of mixing is higher when the
size of the environment is smaller.
This enhancement in mixing time can be viewed as a quantum speedup,
and the mixing time converges to the classical value as the size of
the environment approaches infinity.
When the size of the environment is finite, the duration of the mixing process
is also finite. After the state has converged to a certain
distance from the maximally mixed state, it ceases to approach the
mixed state and saturates to an almost constant distance.
The mean distance (and fluctuations around the mean) in this stabilized
regime of time evolution also depend on the size of the environment,
the distance becoming smaller as the size of the environment increases.
The mean distance was found to follow a power law as a function of
$d_B/d_S$, the ratio between the sizes of the bath and the system.

These results are in contrast with earlier results on mixing processes in
quantum walks,\cite{kt2003,km2008,richter2007} where the walker was
taken to be subject to random uncorrelated noise events.
Those models assumed implicitly that the environment is Markovian,
while the unitary dynamics employed in this
study preserves the memory of the environment.
It has been shown here that when the size of the environment is finite,
the memory can induce qualitatively different behaviour in the long time limit,
and the state of the system tends to fluctuate around an equilibrium state
which is not the maximally mixed state.

The dynamics of the mixing process can be understood in light of some recent
results in quantum information theory and quantum thermodynamics.
Linden \emph{et al}\cite{lpsw2009} have shown that in a large unitarily evolving quantum
system, any small subsystem tends to equilibrate eventually to a particular
state, given that the initial state has support over sufficiently many
eigenstates of the Hamiltonian.
In the context of quantum walks, this translates to the requirement that
the initial state $\ket{\Psi(0)}$ should have support over many eigenstates
of the time evolution operator $U$.
This condition was fulfilled by all the cases considered in this article.

The mechanism of equilibration in closed quantum systems is related
to the surprising fact that in a randomly
chosen pure quantum state, most subsystem states are almost maximally
entangled for some bipartitioning of the system.
Hayden \emph{et al}\cite{hlw2006} have proved some rigorous bounds
for the distance between a typical state and the maximally entangled
state, when both subsystems are large.
The key idea is that in unitary evolution the system samples the states in
a uniform way over time, and since most states are highly entangled, the
small subsystem spends most of the time close to the maximally mixed state.

To understand the systems' tendency to be closer to the maximally
mixed state when $d_B$ is large, it is instructive to look at the
average entropy of a quantum subsystem.
Partitioning a randomly chosen pure state in two systems with
dimensions $d_S$\,$\leq$\,$d_B$, the average entropies of the subsystems
are\cite{page1993,fk1994}
$\langle H \rangle = \sum_{k=d_B+1}^{d_Sd_B} 1/k - (d_S-1)/2d_B$.
Keeping the smaller dimension $d_S$ fixed, the average entropy
increases when $d_B$ increases, and the entropy approaches
the maximum value $\langle H \rangle \rightarrow \ln d_S$ as
$d_B \rightarrow \infty$.
A higher entropy means that the system is closer to the maximally
mixed state, and therefore the distance $D_{\omega}$ is smaller
when $d_B$ is large.

This work might pave way for engineering mixing processes with a
high level of control on their properties.
The mixing time is a standard benchmark quantity for many random walk based
algorithms, and understanding the conditions for improving the mixing time
is essential.
The mixing process is also essential for quantum random number generators,
which require that random numbers are drawn from a distribution that is as
uniform as possible.
By carefully tuning the interaction between the sampled system and its
environment, it could become possible to engineer the distribution
of random variables such that they can be guaranteed to be close to
the completely mixed distribution.

It is generally an interesting question to determine the conditions when
the mixing process is induced by the environment.
As discussed above, simply enlarging the coin dimension does not
always lead to mixing behaviour.
It would also be desirable to have analytical expressions for the
equilibrium state and its distance from the maximally mixed
state $\omega$, and for the standard deviation of the fluctuations
around the equilibrium state.
Finally, it would be interesting to consider how often the system exhibits
recurrences, where the distance $D_{\omega}$ jumps close to its
initial value.
As the size of the system and the environment increases, most states
are close to the maximally state\cite{hlw2006}, and product states
such as the initial state $\ket{\Psi(0)}$ are more unlikely,
so it is expected that recurrences are rare in large systems.

\section*{Acknowledgments}

The author would like to thank G. Brennen and P. \'{C}wikli\'{n}ski for
helpful discussions, and acknowledges support by the Alexander von
Humboldt foundation.

\vspace*{-6pt}   


\vspace{5ex}
\noindent
This is a preprint of an article published in {\it Int. J. Quantum Inform.} {\bf 12} (4), 1450021 (2014)\\
DOI: 10.1142/S021974991450021X\\
\copyright\, World Scientific Publishing Company\\
http://www.worldscientific.com/worldscinet/ijqi

\end{document}